\documentclass[journal=apchd5,manuscript=article ]{achemso}
\usepackage[utf8]{inputenc}
\usepackage[T1]{fontenc}
\usepackage{mathptmx}
\usepackage{etoolbox}
\usepackage{xcolor}
\usepackage{graphicx}
\usepackage{dcolumn}
\usepackage{bm}
\usepackage{setspace}
\usepackage{soul}  
\usepackage{lineno}
\usepackage{siunitx} 
\usepackage[version=3]{mhchem} 

\author{Tecla Gabbrielli}
 \affiliation{CNR-INO -- Istituto Nazionale di Ottica, Via Carrara, 1 -- 50019 Sesto Fiorentino FI, Italy}
\alsoaffiliation{LENS -- European Laboratory for Non-Linear Spectroscopy, Via Carrara, 1 -- 50019 Sesto Fiorentino FI, Italy}
\altaffiliation{These authors equally contributed to this work.}
\author{Jacopo Pelini}%
\altaffiliation{These authors equally contributed to this work.}
 \affiliation{CNR-INO -- Istituto Nazionale di Ottica, Via Carrara, 1 -- 50019 Sesto Fiorentino FI, Italy}
\alsoaffiliation{LENS -- European Laboratory for Non-Linear Spectroscopy, Via Carrara, 1 -- 50019 Sesto Fiorentino FI, Italy}
\author{Chenhong Zhang}
 \affiliation{CNR-INO -- Istituto Nazionale di Ottica, Via Carrara, 1 -- 50019 Sesto Fiorentino FI, Italy}
\alsoaffiliation{LENS -- European Laboratory for Non-Linear Spectroscopy, Via Carrara, 1 -- 50019 Sesto Fiorentino FI, Italy}
\author{Francesco Cappelli}
\affiliation{CNR-INO -- Istituto Nazionale di Ottica, Via Carrara, 1 -- 50019 Sesto Fiorentino FI, Italy}
\alsoaffiliation{LENS -- European Laboratory for Non-Linear Spectroscopy, Via Carrara, 1 -- 50019 Sesto Fiorentino FI, Italy}
\author{Mario Siciliani de Cumis}
\affiliation{ASI, Italian Space Agency, Località Terlecchia, Matera, 75100, Italy}
\alsoaffiliation{CNR-INO -- Istituto Nazionale di Ottica, Via Carrara, 1 -- 50019 Sesto Fiorentino FI, Italy}
\email{mario.sicilianidecumis@asi.it}
\author{Stefano Dello Russo}
\affiliation{ASI, Italian Space Agency, Località Terlecchia, Matera, 75100, Italy}
\author{Maria Concetta Canino}
\affiliation{CNR-ISMN, The Institute for the Study of Nanostructured Materials of the National Research Council of Italy, Via P. Gobetti 101, Bologna, 40129, Italy}
\author{Alberto Roncaglia}
\affiliation{CNR-ISMN, The Institute for the Study of Nanostructured Materials of the National Research Council of Italy, Via P. Gobetti 101, Bologna, 40129, Italy}
\author{Paolo De Natale}
\affiliation{CNR-INO -- Istituto Nazionale di Ottica, Via Carrara, 1 -- 50019 Sesto Fiorentino FI, Italy}
\alsoaffiliation{LENS -- European Laboratory for Non-Linear Spectroscopy, Via Carrara, 1 -- 50019 Sesto Fiorentino FI, Italy}
\author{Simone Borri}
\affiliation{CNR-INO -- Istituto Nazionale di Ottica, Via Carrara, 1 -- 50019 Sesto Fiorentino FI, Italy}
\alsoaffiliation{LENS -- European Laboratory for Non-Linear Spectroscopy, Via Carrara, 1 -- 50019 Sesto Fiorentino FI, Italy}

\title
  {Self-mixing-based photoacoustic sensing}
\begin{document}
\singlespacing




\begin{abstract}
Versatile, ultracompact, easy-to-handle, high-sensitivity sensors are compelling tools for in situ pivotal applications, such as medical diagnostics, security and safety assessments, and environmental control. In this work, we combine photoacoustic spectroscopy and feedback interferometry, proposing a novel trace-gas sensor equipped with a self-mixing readout. This scheme demonstrates a readout sensitivity comparable to that of bulkier state-of-the-art balanced Michelson-interferometric schemes, achieving the same spectroscopic performance in terms of signal-to-noise ratio (SNR) and minimum detection limit (MDL). At the same time, the self-mixing readout benefits from a reduced size and a lower baseline, paving the way for future system downsizing and integration while offering a higher detectability for lower gas concentrations. Moreover, the intrinsic wavelength independence of both self-mixing and photoacoustic techniques allows the applicability and tailorability of the sensor to any desired spectral range.
\end{abstract}
\newpage


Trace-gas sensing is at the basis of many key applications in contemporary society. Spanning from environmental monitoring~\cite{dhall2021review,saxena2023review} and medical diagnostics~\cite{banik2020exhaled,kaloumenou2022breath}, to industrial process~\cite{jiang2020review,lewen2022sensitive} and safety \& security control~\cite{vasiliev2013sensors,to2020recent}, all these pivotal applications demand compact, reliable, high-sensitivity, energy-consumption-effective, and versatile trace-gas sensors to monitor specific, application-driven gas species, such as \ce{CO2}, \ce{N2O}, \ce{CH4}, \ce{CO} and \ce{H2F}, to mention a few. Many of the cited gases are characterized by strong rovibrational transitions in the mid-infrared (MIR), which are at the basis of high detection sensitivity. At these wavelengths, cavity-ring-down spectroscopy techniques have demonstrated sensitivities of up to the part-per-quadrillion (ppq) level~\cite{galli2016spectroscopic,jiang2024mid,duddy2025beyond}, but at the cost of a bulky and complex setup, which is unable to meet the demands of portability and compactness. A good alternative is offered by the Photoacoustic (PA) technique, which ensures high and competitive detection sensitivity levels, down to the sub-part-per-trillion~\cite{tomberg2018,wang2022doubly,nie2024sub}, with the advantage of higher compactness and versatility. The PA is a wavelength-independent laser spectroscopic technique based on the detection of the acoustic wave emitted by the target molecules relaxing via a non-radiative channel~\cite{rosencwaig1980photoacoustic}. It allows for easy-to-handle setups consisting only of an excitation source, a chamber filled with a mixture containing the target trace gas, and an acoustic detector coupled with a signal readout. Based on the chosen detector, the signal readout can be optical or electrical.

Electrical readouts are typically deployed in combination with detectors such as quartz tuning forks~\cite{kosterev2002quartz, patimisco2014quartz} or conventional microphones~\cite{frederiksen2008condenser}, and they have the advantage of easy miniaturization depending only on the electronic architecture~\cite{li2022compact}.
Optical readout, such as the one based on Michelson interferometers and silicon-based cantilevers~\cite{wilcken2003optimization,pelini2024new,Pelini2025}, can offer, in principle, higher sensitivity when compared to the electrical one, but at the cost of a bulkier experimental implementation. A possible solution aimed at scaling down the size is provided by optical fiber-based readout~\cite{chen2018ultra,zhao2025fiber}
A different solution providing potential integrability can be offered by the self-mixing (SM) interferometric technique, yet unexplored photoacoustic-based sensors for acoustic wave transduction. In SM interferometry, the probe source (typically, a semiconductor laser) is also the detector, allowing for a very compact optical readout design~\cite{Giuliani:2002, leng2011demonstration, Vezio:2025, Ottomaniello:19}. The optical signal of the probe laser is back-reflected from the target (e.g., a membrane) and re-injected into its waveguide. Here, the back-reflected signal interferes with the intra-cavity radiation with a phase shift dependent on the target position, determining a change in the laser working point. This allows, therefore, the measurement of the target information (e.g., displacement or oscillation frequency of the membrane) via monitoring the voltage drop at the laser terminals, without any need for an additional optical and bulkier detection system~\cite{Giuliani:2002,gabbrielli2025selfmixarxiv}. Since the probe laser can be placed in proximity to the target, the whole detection system (transducer plus optical readout) can be made extremely compact and, potentially, integrated into a chip-scale module.  In this work, we demonstrate that SM readout, when applied to a cantilever-based photoacoustic sensor, ensures the same sensitivity levels as standard and bulkier tabletop-sized interferometric readouts, but in a tightly packed design and with a much reduced number of implemented optics.  

\begin{figure*}[hbt!] 
\centering\includegraphics[width=0.85\textwidth]{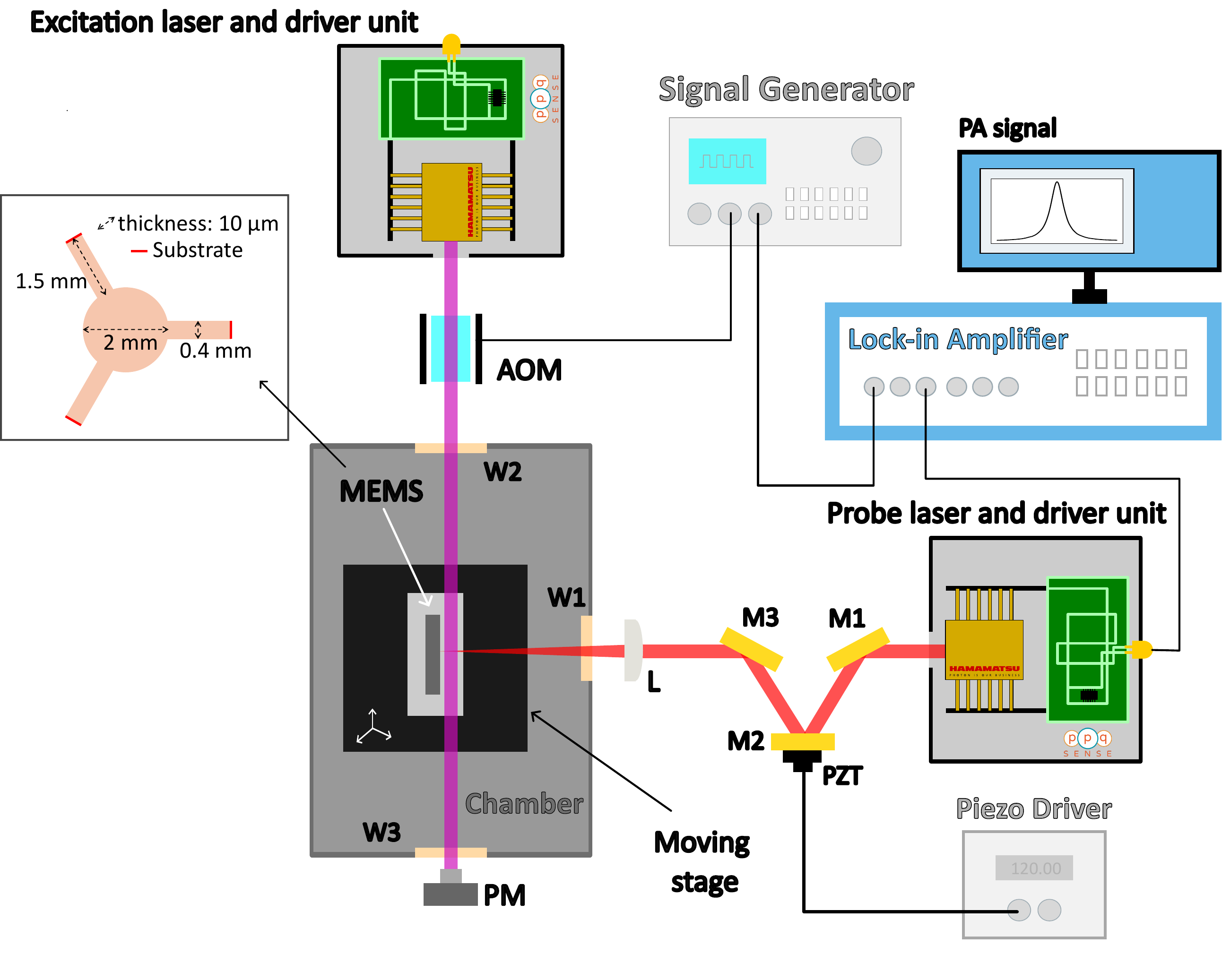}
\caption{Sketch of the self-mixing-based photoacoustic setup. In the figure, W1, W2 and W3 are ZnSe windows; PM is a power meter; M1, M2 and M3 are gold-coated mirrors. The piezo-electric transducer (PZT) mounted on the M2 mirror is used to optimize the phase match condition and, therefore, the amplitude of the self-mixing readout signal. \textbf{Inset}: Zoom of the 3-armed spring MEMS.}
\label{fig:setup}
\end{figure*}

The sensor design is shown in Fig.~\ref{fig:setup}. The gas sample is loaded into a portable chamber (l$\times$w$\times$h: 140mm$\times$250mm$\times$110mm) equipped with a 3-armed spring Micro-Electro-Mechanical System (MEMS) mounted on a 5-axis translating stage for alignment purposes. On the shorter side, the excitation beam propagates inside the chamber, entering through a ZnSe window (W2). After passing in front of the MEMS, the beam exits the cell on the opposite side, where it is monitored via a power meter. The excitation laser used is a commercial Distributed Feedback Quantum Cascade Laser (DFB QCL) by Hamamatsu Photonics emitting around \SI{4.57}{\micro \meter} at room temperature. The working conditions of the excitation beam (i.e., optical power and wavelength controlled via temperature and bias current) are set to detect the P(36) absorption line of \ce{N2O} centered at 2189.273 $\mathrm{cm^{-1}}$. An acousto-optic modulator (AOM) provides a periodic excitation beam amplitude modulation at the MEMS resonance frequency, allowing for an on-resonance MEMS oscillation driven by non-radiative relaxation of the target molecules. The MEMS (inset Fig.~\ref{fig:setup}) is investigated in its fundamental resonance mode, whose measured frequency falls in the range (5.2-5.3) kHz, depending on the gas pressure. The MEMS oscillation is detected via a self-mixing interferometric readout. To this extent, a probe laser propagates back and forth in the cell via a ZnSe window (W1), back-reflected by the MEMS. The back-reflected signal is then injected into the laser waveguide, where it interferes with the intra-cavity radiation, generating the self-mixing signal. The latter can be detected directly by monitoring the laser bias current \cite{Vezio:2025,leng2011demonstration, Giuliani:2002}. In this configuration, no extra photodetectors are required, with the advantage of an easy scale-down in the whole sensor dimension. The photoacoustic signal is then reconstructed by demodulating in first-harmonic the self-mixing signal with a lock-in amplifier and using the modulation signal sent to the AOM as a reference. In this work, the probe laser used for the self-mixing readout is another commercial DFB QCL by Hamamatsu Photonics emitting around \SI{4.57}{\micro \meter} at room temperature (i.e., a twin of the excitation laser). However, we remark that both the photoacoustic technique and the self-mixing readout are wavelength-independent, and, in principle, this sensor can be tailored to any color wavelength on both excitation and probe sides. 

During the measurement process, the chamber was filled with a mixture made by the target gas, i.e. \ce{N2O}, diluted at a trace level of \SI{200}{ppm} within another buffer gas, \ce{N2} in our case. The excitation laser bias current was scanned via a slow triangular modulation (f = \SI{4}{mHz}) by means of the driver unit around the target \ce{N2O} absorption line. Being the lock-in demodulation fixed at the MEMS resonance frequency, the acquisition of the SM demodulated signal over time allowed us to retrieve the entire \ce{N2O} photoacoustic absorption spectrum. Then, this latter signal can be analyzed to extract information about the detection sensitivity performance (i.e., minimum detection limit) of the developed system. This procedure was repeated for several sample pressures within \SI{5} and \SI{200} mbar. During all the measurements, the SM probe laser was operated at a fixed working condition of temperature and bias current far from both threshold and saturation. Before measuring the entire PA absorption spectrum, a preliminary signal optimization was carried out at each working pressure. At this stage, the current of the excitation laser beam was set at the \ce{N2O} absorption peak, and the AOM modulation frequency was chosen to match the MEMS resonance frequency. Then, to maximize the phase matching between the incident and the back-reflected beams of the SM probe laser, the position of the mirror M2 was manually optimized by changing the driving DC voltage applied to a piezo actuator (PTZ).

\begin{figure*}[ht!]
\centering\includegraphics[width=0.95\textwidth]{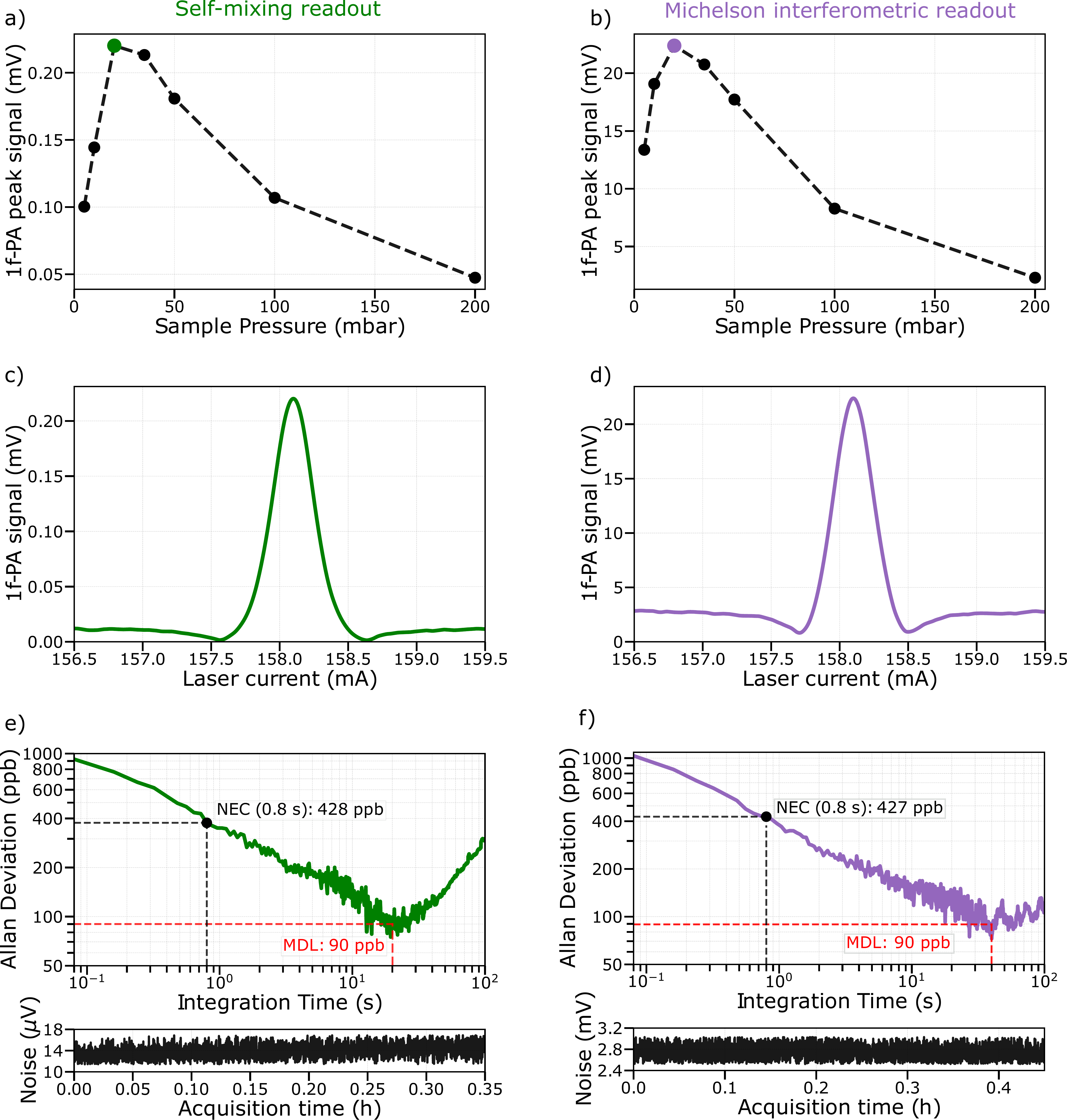}
\caption{Comparison between the self-mixing and the Michelson-interferometric readouts. \textbf{(a)} photoacoustic peak signal as a function of the sample pressure, \textbf{(c)} PA signal at the best working pressure, and \textbf{(e)} Allan-Werle deviation analysis for the self-mixing-based readout. \textbf{(b)} photoacoustic peak signal as a function of the sample pressure, \textbf{(d)} PA signal at the best working pressure, and \textbf{(f)} Allan-Werle deviation analysis for the Michelson-interferometric-based readout. 
The y-axis of the Allan-Werle plots is converted from mV to gas noise equivalent concentration (i.e., ppb) via the relation $c/SNR$, where $c$ is the fixed trace-gas concentration within the sample, and SNR is the ratio between the PA peak signal and the noise level.}
\label{Comparison}
\end{figure*}

For a fair performance evaluation of the developed SM-based photoacoustic readout, we compare its experimental performance with the one obtained with the same PA excitation scheme but coupled with a state-of-the-art Michelson interferometric readout~\cite{Pelini2025}. In this latter case, the readout probe (i.e., an HeNe laser), is split into two identical paths: one, being the reference arm, impinges on a fixed mirror, and the other is focused on the oscillating 3-arm spring MEMS. The two back-reflected beams are collected at a beam splitter and sent to a balanced photo-detector, and the balanced interferometric output is demodulated with the strategy previously introduced. Compared with the SM-based readout, the Michelson interferometer requires a higher level of complexity, namely a higher number of optics and a complex polarization control system.

Fig.~\ref{Comparison} summarizes the results of the comparison performed under the same experimental conditions (i.e., gas sample pressures, trace-gas concentration, and excitation laser wavelength and power) and measurement parameters (i.e., AOM modulation signal, ramp applied to scan the excitation laser wavelength, and lock-in settings such as bandwidth, fixed at \SI{0.1}{Hz}, corresponding to an integration time of \SI{0.8}{s}). According to Fig.~\ref{Comparison}, the trend of the PA peak signal as a function of the sample pressure is substantially the same for the two readout schemes (Fig.~\ref{Comparison}~\textbf{(a)} and~\ref{Comparison}~\textbf{(b)}). Specifically, after increasing at low pressure, the PA signal reaches its maximum at \SI{20}{mbar}, and then it decreases at higher pressure, following the MEMS quality factor degradation. Fig.~\ref{Comparison}~\textbf{(c)} and Fig.~\ref{Comparison}~\textbf{(d)} show the PA spectra acquired with the two readout setups at the best working point (\SI{20}{mbar}). In the case of self-mixing readout, both the PA signal and the noise are scaled down by approximately two orders of magnitude with respect to the Michelson readout. 
This leads to a comparable SNR and, consequently, comparable Noise Equivalent Concentration (NEC) values, i.e., \SI{428}{ppb} and \SI{427}{ppb} for the self-mixing and balanced interferometric readouts, respectively. This demonstrates that the two readout schemes can be equally used, achieving the same final detection sensitivity, and that the sensor performance is substantially limited by the MEMS intrinsic noise. Furthermore, it is worth noting that in the case of self-mixing readout, the acquired PA spectrum is characterized by a lower baseline, which potentially enables higher sensitivity at low gas concentration levels, thus potentially extending the dynamic range of the sensor. 

Additionally, long-term noise traces have been acquired at \SI{20}{mbar} with both systems by moving the laser at a higher operating current to avoid residual interaction between the excitation beam and the target gas. Both the traces have been acquired by setting the lock-in bandwidth to \SI{1}{Hz}, corresponding to an integration time of \SI{0.08}{s}. With these traces, Allan-Werle deviation analysis has been performed to evaluate the achievable MDLs \cite{giglio2015allan}. According to the results of this analysis, both readout schemes can achieve an MDL of \SI{90}{ppb} with integration times of \SI{20}{s} (Fig. \ref{Comparison} \textbf{(e)}) and \SI{40}{s} (Fig. \ref{Comparison} \textbf{(f)}). This evidence clearly highlights that a compact and easy-to-handle self-mixing readout can be used in place of the bulkier Michelson interferometer without losing sensitivity. 

To summarize, in this work, we have presented a novel PA sensor embedded with a self-mixing readout. When compared to a more standard Michelson interferometric readout, SM shows the same MDL sensitivity level and signal-to-noise ratio performances, with the advantage of a more compact and easier-to-handle solution in view of chip-scale sensor implementation and a potentially higher dynamic range. Thanks to the wavelength independence of both the PA technique and self-mixing interferometry, such a sensor can be tailored for operation at the desired wavelength according to the requirements of the specific application. 

\paragraph{Data Availability Statement}
The data that support the findings of this study are available from the
corresponding author upon reasonable request.

\begin{acknowledgement}
The authors gratefully acknowledge PpqSense S.r.l. and Hamamatsu Photonics for providing the driver units (QubeCLs) and the lasers, respectively. 
The authors acknowledge financial support from the
European Union with the Next Generation EU with the Italian National Recovery and Resilience Plan (NRRP), Mission 4, Component 2, Investment 1.3, CUP D43C22003080001, partnership on “Telecommunications of the Future” (PE00000001—program “RESTART”) and the I-PHOQS Infrastructure ”Integrated infrastructure initiative in Photonic and Quantum Sciences” [IR0000016, ID D2B8D520], with the Laserlab-Europe Project [G.A. n.871124], with the MUQUABIS Project “Multiscale quantum bio-imaging and spectroscopy” [G.A. n.101070546], with the ForeSight Project (GA n.101168521 ), with the QUID project "Quantum Italy Deployment" [G.A. No 101091408]; from the Italian ESFRI Roadmap (Extreme Light Infrastructure - ELI Project); from ASI and CNR under the Joint Project “Laboratori congiunti ASI-CNR nel settore delle Quantum Technologies (QASINO)” (Accordo Attuativo n. 2023-47-HH.0); from the Italian Ministero dell'Università e della Ricerca (project PRIN-2022KH2KMT QUAQK). The project QOSTRAD "Quantum-enhanced optomechanical sensors for trace gas detection" (ID 59182/2024) funded by the European Union - Next Generation EU (PNRR-MUR) PE0000023 - NQSTI (National Quantum Science and Technology Institute).

\end{acknowledgement}

\bibliography{ref}

\end{document}